\documentclass[final,5p,times,twocolumn]{elsarticle}
\usepackage{booktabs}
\usepackage{siunitx}
\usepackage{lineno}
\usepackage{amsmath,amssymb}
\usepackage{graphicx}
\usepackage{multirow}
\usepackage{microtype}
\biboptions{sort&compress}
\usepackage{xurl}
\usepackage[hidelinks]{hyperref}
\newcommand{\Kbar}{\bar{K}}
\newcommand{\KNN}{\Kbar NN}

\newcommand{\gev}{\mathrm{GeV}}
\newcommand{\mev}{\mathrm{MeV}}
\newcommand{\cunit}{c}
\journal{Physics Letters B}

\begin{document}

\begin{frontmatter}

\title{Evidence for a \texorpdfstring{$\bar{K}NN$}{KbarNN} quasi-bound state in the
\texorpdfstring{$\gamma d \to K^0 \Lambda p$}{gamma d -> K0 Lambda p} reaction}
\author[rcnp,kyoto]{R.~Kobayakawa}
\author[korea]{J.~K.~Ahn}
\author[rcnp]{S.~Ajimura}
\author[academia-sinica]{W.-C.~Chang}
\author[academia-sinica]{M.-L.~Chu}
\author[rcnp]{S.~Dat{\'e}}
\author[kyoto]{T.~Gogami}
\author[rcnp]{T.~A.~Hashimoto}
\author[rcnp]{T.~Hotta}
\author[rcnp]{T.~Ishikawa}
\author[rcnp]{H.~Katsuragawa}
\author[rcnp,nagoya]{H.~Kohri}
\author[riken]{Y.~Ma}
\author[raris]{M.~Miyabe}
\author[rcnp]{K.~Mizutani}
\author[imp]{N.~Muramatsu}
\author[rcnp]{T.~Nakano}
\author[dalat]{T.~H.~Nam}
\author[kyoto-sangyo]{M.~Niiyama}
\author[tokyo]{Y.~Nozawa}
\author[rcnp]{Y.~Ohashi}
\author[rcnp]{H.~Ohkuma\fnref{deceased}}
\author[raris]{H.~Ohnishi}
\author[tokyo]{T.~Ohta}
\author[saskatchewan]{C.~Rangacharyulu}
\author[rcnp]{S.~Y.~Ryu}
\author[raris]{Y.~Sada}
\author[raris]{H.~Shimizu}
\author[gifu,rcnp]{M.~Sumihama}
\author[rcnp]{S.~Suzuki}
\author[rcnp]{S.~Tanaka}
\author[raris]{A.~O.~Tokiyasu}
\author[kyoto]{N.~Tomida}
\author[rcnp]{K.~Watanabe}
\author[rcnp]{T.~Yorita}
\author[kek]{C.~Yoshida}
\author[rcnp]{M.~Yosoi}
\author[]{\\ (LEPS2/Solenoid Collaboration)}

\address[rcnp]{Research Center for Nuclear Physics, The University of Osaka, Ibaraki, Osaka 567-0047, Japan}
\address[korea]{Department of Physics, Korea University, Seoul 02841, Republic of Korea}
\address[academia-sinica]{Institute of Physics, Academia Sinica, Taipei 11529, Taiwan}
\address[kyoto]{Department of Physics, Kyoto University, Kyoto 606-8502, Japan}
\address[riken]{RIKEN Nishina Center for Accelerator-Based Science, Wako, Saitama 351-0198, Japan}
\address[raris]{Research Center for Accelerator and Radioisotope Science (RARiS), Tohoku University, Sendai, Miyagi 982-0826, Japan}
\address[imp]{Institute of Modern Physics, Chinese Academy of Sciences, Lanzhou 730000, China}
\address[dalat]{Dalat Nuclear Research Institute, Dalat, Lam Dong, Vietnam}
\address[kyoto-sangyo]{Department of Physics, Kyoto Sangyo University, Kyoto 603-8555, Japan}
\address[tokyo]{Department of Radiology, The University of Tokyo Hospital, Tokyo 113-8655, Japan}
\address[saskatchewan]{Department of Physics and Engineering Physics, University of Saskatchewan, Saskatoon, Canada SK S7N 5E2}
\address[gifu]{Department of Education, Gifu University, Gifu 501-1193, Japan}
\address[kek]{Institute of Particle and Nuclear Studies, High Energy Accelerator Research Organization (KEK), Tsukuba, Ibaraki 305-0801, Japan}
\address[nagoya]{Nagoya University, Furocho, Chikusa, Nagoya, 464-8602, Japan}
\fntext[deceased]{Deceased}

\begin{abstract}
The $\gamma d \to K^{0}\Lambda p$ reaction has been studied to search for a $\bar{K}NN$ quasi-bound state
using the LEPS2 solenoid spectrometer at SPring-8.
A localized enhancement concentrated at low $q$ is observed in the $(M_{\Lambda p},q)$ distribution below the $K^-pp$ mass threshold,
where $M_{\Lambda p}$ is the $\Lambda p$ invariant mass and $q \equiv |\vec{p}_{\gamma}-\vec{p}_{K^{0}}| = |\vec{p}_{\Lambda p}|$ is the momentum transfer.
A two-dimensional $(M_{\Lambda p},q)$ fit demonstrates that the enhancement near the threshold is statistically significant, yielding a local significance of $7.3\,\sigma$.
The enhancement is characterized by effective shape parameters including the Breit--Wigner mass $M$ and width $\Gamma$, and a Gaussian form-factor momentum scale $Q$:
$M = 2.354 \pm 0.011(\mathrm{stat.})^{+0.009}_{-0.005}(\mathrm{syst.})~\gev/\cunit^{2}$,
$\Gamma = 0.055 \pm 0.023(\mathrm{stat.})^{+0.031}_{-0.009}(\mathrm{syst.})~\gev/\cunit^{2}$, and
$Q = 0.350 \pm 0.041(\mathrm{stat.})^{+0.035}_{-0.010}(\mathrm{syst.})~\gev/\cunit$.
The obtained results support the existence of a $\bar{K}NN$ quasi-bound state in photoproduction.
\end{abstract}


\end{frontmatter}

\section{Introduction}
A fundamental question in hadron and nuclear physics is how nuclear structure is dynamically modified when a meson is embedded in the system.
Among various meson--nucleon interactions, the antikaon--nucleon ($\bar{K}N$) interaction is of particular interest because it exhibits a strong attraction in the isospin-$I=0$ channel, a feature closely connected to the $\Lambda(1405)$ resonance located below the $\bar{K}N$ threshold~\cite{HyodoJido2012}.
This strong attraction may overcome the conventional nuclear repulsive forces, suggesting the possible formation of highly compact kaonic nuclei and offering a pathway to explore high-density matter~\cite{Akaishi:2002Kpp}.
The simplest prototype of such highly compact kaonic nuclei is the $\bar{K}NN$ state, which serves as the minimal laboratory for exploring the competition between the dominant $\bar{K}N$ attraction and the conventional short-range $NN$ repulsion. This balance defines the unique spatial compactness of the $\bar{K}NN$ system~\cite{Gal2013,HyodoWeise2022}.
Theoretical calculations predict a quasi-bound state with a binding energy of about $10$--$90~\mev$ and a decay width of $40$--$100~\mev$, depending on the underlying interaction model~\cite{Tolos2020, Shevchenko:2007Kpp, Dote:2009Kpp}.
The ground state of the $\bar{K}NN$ system is expected to have $J^{P}=0^{-}$, because the system favors an $I=0$ $\bar{K}N$ pair to maximize the strong attraction, which in turn places the two nucleons in a relative $S$-wave, isospin-$I=1$, spin-singlet ($S=0$) configuration under the Pauli principle~\cite{HyodoWeise2022, Akaishi:2002Kpp}.

Experimental searches for the $\bar{K}NN$ state are complicated by ambiguities associated with reaction mechanisms and final-state interactions~\cite{Iwasaki2023}.
While several hadron-beam experiments report subthreshold enhancements interpreted as possible $\bar{K}NN$ candidates~\cite{Agnello:2005Kpp,Yamazaki:2010Kpp,Ichikawa:2015PTEP}, other measurements do not show any significant structure, and conventional explanations such as multi-nucleon absorption, quasi-free hyperon production, reflection from $N^*$ resonances, or a threshold cusp effect have also been proposed~\cite{Tokiyasu2014,Agakishiev:2015HADES,Hashimoto:2015Kpp,Iwasaki2023}.
Thus, evidence for the $\bar{K}NN$ state remains inconclusive.
A major step forward was made by the J-PARC E15 experiment~\cite{Sada:2016Kpp,Yamaga2020}, which observed a pronounced structure consistent with a $\bar{K}NN$ quasi-bound state in the exclusive $K^{-}{}^{3}\mathrm{He} \to \Lambda p n$ reaction.
However, despite the strong theoretical motivation for a $J^P = 0^-$ state, the spin and parity of the observed $\bar{K}NN$ candidate remain to be determined experimentally.

To clarify the nature of the $\bar{K}NN$ state, it is important to study the system through a different entrance channel.
Photon-induced reactions on the deuteron provide a complementary approach.
In the present case, the deuteron target is in a spin-triplet configuration, whereas the $\bar{K}NN$ state is expected to contain a spin-singlet nucleon pair; photoproduction can in principle populate spin-singlet configurations through spin-flip amplitudes.
Experimentally, the formation of the quasi-bound state is expected to be enhanced at low momentum transfer, which kinematically corresponds to forward kaon emission.
For such forward production, $t$-channel exchange processes are expected to contribute significantly~\cite{Sumihama:2006PRC, Hicks:2009PRL, Scheluchin:2022PLB}.
For the $K^{0}$ photoproduction channel, the $t$-channel $K^0$ exchange is absent, so other mechanisms, such as $K^{*}$ exchange and resonance contributions, can become relatively more important~\cite{Mart:2011}.
Production may also receive contributions from a $\Lambda(1405)$ doorway mechanism, especially in the forward kaon region~\cite{Yamazaki:2007Kpp}.
These considerations motivate an exclusive photoproduction study on the deuteron as an independent test of the kaonic-nucleus picture.

An earlier inclusive search in the $\gamma d \to K^{+}\pi^{-}X$ reaction by the LEPS Collaboration found no significant signal, mainly because of large quasi-free background contributions~\cite{Tokiyasu2014}.
In this Letter, we report evidence for a $\bar{K}NN$ quasi-bound state in the $\gamma d \to K^{0}\Lambda p$ reaction for the first time with the LEPS2 spectrometer~\cite{LEPS2}.
\section{Experimental setup and data sample}

The experiment was performed using the LEPS2 beamline~\cite{Muramatsu:2021bpl} at SPring-8.
A multi-GeV photon beam is produced by backward Compton scattering of linearly-polarized laser light with a wavelength of 355 nm off the circulating electrons in the 8~GeV electron storage ring, yielding linearly-polarized photons with energies below 2.4 GeV.
The energy of each produced photon above 1.3 GeV is determined event-by-event by measuring the momentum of the recoil electron in a photon-tagging system, with an energy resolution of approximately 12~MeV.
The data were collected using a large-acceptance solenoid spectrometer~\cite{LEPS2, Niiyama:2024LEPS2}.
The spectrometer is based on a large solenoid magnet with a 0.9~T central field, in which all the detectors except for the photon-tagging system are installed.
It covers large angles in the laboratory frame, being optimized for detection of final-state multi-particles.
The tagged photon beam was incident on a 15-cm-long liquid-deuterium (LD$_2$) target.
The integrated number of tagged photons used in this analysis is $3.6\times10^{12}$.
The momentum scale and beam energy have been confirmed to be accurate through the separate analyses of known reactions, yielding masses consistent with the Particle Data Group values.

The detector system comprises an inner plastic scintillator array surrounding the target, a time projection chamber (TPC) for charged-particle tracking, a barrel-shaped resistive plate chamber (BRPC)~\cite{Watanabe:2018cod} for time-of-flight measurements, and a barrel lead-scintillator sampling electromagnetic calorimeter.
The inner scintillator array serves as a start counter for trigger generation and consists of a hexagonal forward sheet and six side sheets.
The TPC serves as the primary tracking detector, while the BRPC provides time-of-flight information used for particle identification in combination with the energy loss measured in the TPC.
Charged tracks are reconstructed primarily in the TPC acceptance ($40^{\circ}\lesssim \theta_\mathrm{lab}\lesssim110^{\circ}$).
Owing to the material budget inside the TPC (target cell, inner scintillator array, and inner electric-field cage for TPC), protons with momenta below $\sim0.27$~GeV/$c$ are not reconstructed, which suppresses quasi-free $\gamma n\to K^{0}\Lambda$ events with a low-momentum spectator while retaining the exclusive $\gamma d\to K^{0}\Lambda p$ signal.
The remaining quasi-free contribution is dominated by the high-momentum tail of the Fermi motion inside the deuteron.
Details of the data-acquisition system are described in Ref.~\cite{Ryu:2025cjn}.
The data acquisition efficiency was maintained at approximately 85\% with a typical trigger rate of $\sim$120~Hz and a photon tagging rate of 1.5~MHz.
Signal and background processes are studied using a Geant4-based Monte Carlo (MC) simulation~\cite{Allison2016Geant4, Agostinelli2003Geant4, Allison2006Geant4}.

Events were collected with a trigger condition based on the inner scintillator array, requiring hits in at least two side counters in coincidence with a signal from the photon tagging system.
To suppress background events not originating from the target, events with hits in an upstream veto counter and a downstream $e^{+}e^{-}$ veto counter were rejected online.
The upstream veto removes charged particles contaminating the incident photon beam, while the downstream veto counter rejects $e^{+}e^{-}$ pairs generated in the target area by electromagnetically converted photons.

\section{Event reconstruction and selection}
\subsection{Particle identification and event selection}
In the offline analysis, we selected events containing two proton candidates and one $\pi^{-}$ candidate.
Events with multiple recoil-electron tracks in the photon tagging system were rejected to ensure an unambiguous assignment of the incident-photon energy.
As a track-quality requirement, each reconstructed track was required to have at least 12 hits in the TPC.
Charged-particle momenta were determined from the track curvature in the magnetic field, with energy-loss corrections through the detector materials.
For particle identification, we employed a likelihood method combining the energy deposit per unit length in the TPC and time-of-flight information from the BRPC.
For each charged track, the likelihood for each particle hypothesis $i$ out of $\pi$, $K$, $p$, and $d$ was constructed as
\begin{equation}
\mathcal{L}_{i}=\mathcal{P}_\mathrm{TPC}(dE/dx\,|\,i)\times \mathcal{P}_\mathrm{BRPC}({\rm TOF}\,|\,i),
\end{equation}
where $\mathcal{P}_\mathrm{TPC}$ and $\mathcal{P}_\mathrm{BRPC}$ are normalized probability density functions.
We defined the relative likelihood $R_{p}\equiv \mathcal{L}_{p}/\sum_{i}\mathcal{L}_{i}$ and selected proton candidates with $R_{p}>0.1$.
The $\pi^{-}$ candidate was identified solely by its negative charge, as the requirement of two identified protons combined with the reaction kinematics suppresses the possibility of misidentification of $K^-$ as $\pi^-$.

Because the final state contains two protons, the daughter proton from the $\Lambda$ decay was identified in the following way.
For each event, we evaluated a likelihood for each $\pi^{-}p$ pairing using the $\pi^- p$ invariant mass and the distance of closest approach (DCA) between the $\pi^{-}$ and proton tracks:
\begin{equation}
\mathcal{L}=\mathcal{P}_\mathrm{inv}(M_{\pi^- p})\times \mathcal{P}_\mathrm{DCA}(\mathrm{DCA}_{\pi^- p}),
\end{equation}
where $\mathcal{P}_\mathrm{inv}$ was modeled as a Gaussian centered at the $\Lambda$ centroid mass and $\mathcal{P}_\mathrm{DCA}$ followed an exponential form of the distance.
Among the two proton candidates, the proton that maximized $\mathcal{L}$ was assigned as the daughter proton from the $\Lambda$ decay.
After this assignment, we reconstructed both the $\Lambda$-decay vertex and the $\Lambda p$-production vertex.

To isolate the exclusive $\gamma d\to K^{0}\Lambda p$ channel, we first required the absence of hits in the forward scintillator counter, ensuring that no forward-going charged particle was present.
This requirement was consistent with the large-angle acceptance of the spectrometer and suppressed background contributions accompanied by forward charged particles.
To reduce contamination from channels with additional neutral particles (mainly photons), we rejected events with a neutral cluster in the electromagnetic calorimeter that was not attributed to any charged track reconstructed by the TPC and BRPC.

We further required a displaced $\Lambda$-decay vertex with a flight length $>18$~mm (measured as the distance between the production and decay vertices) and a small DCA between the $\pi^{-}$ track and the proton from the $\Lambda$ decay, $\mathrm{DCA}<6$~mm.
Mass selections were applied using the reconstructed $\Lambda$-invariant mass and the missing mass $\mathrm{MM}(\pi^-pp)$ for the $K^{0}$ hypothesis:
$|M_{p\pi^{-}}-M_{\Lambda}|<10~\mev/\cunit^{2}$ and $|\mathrm{MM}(\pi^{-}pp)-M_{K^{0}}|<250~\mev/\cunit^{2}$,
where $\mathrm{MM}(\pi^- pp)$ is the missing mass of the $\gamma d \to \pi^- pp X$ reaction derived from the four-momenta of the incident photon, the target deuteron, and the three detected charged particles.
The $\Lambda$-sideband regions used to model the non-strange background were defined symmetrically relative to the $\Lambda$ centroid by the mass difference $15 < |M_{p\pi^{-}}-M_{\Lambda}| < 25~\mev/\cunit^{2}$.
\begin{figure}[htb]
  \centering
  \includegraphics[width=\linewidth]{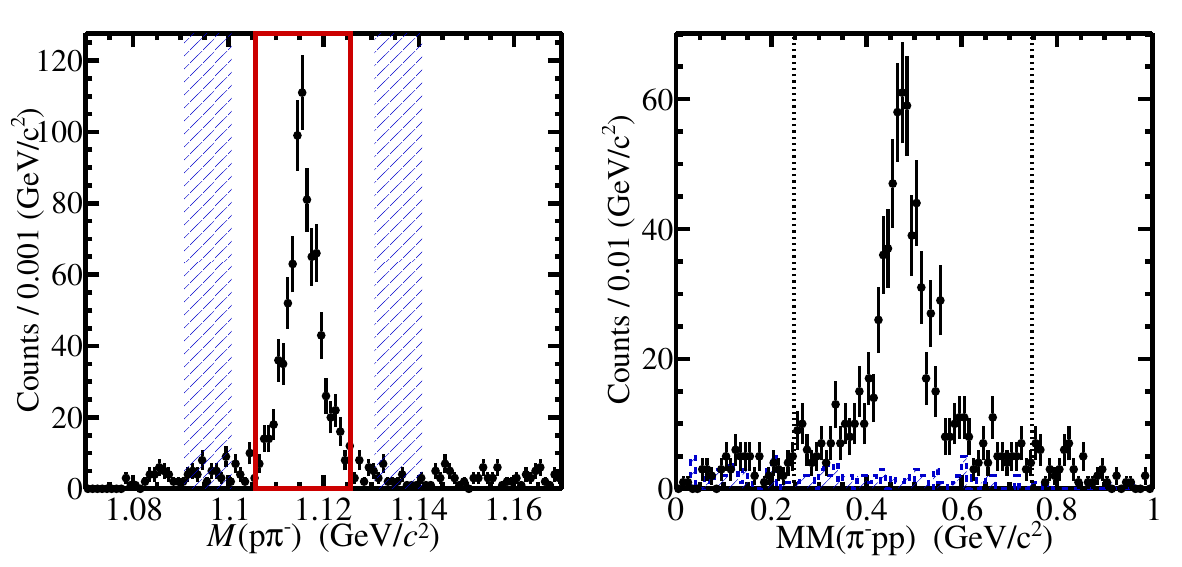}
  \caption{
The $p\pi^-$ invariant-mass and $\pi^-pp$ missing-mass distributions for the exclusive $\gamma d \to K^{0}\Lambda p$ candidates.
(Left) The invariant mass $M_{p\pi^-}$ after requiring the $K^{0}$ missing-mass window,
$\left|\mathrm{MM}(\pi^-pp)-M_{K^{0}}\right|<250~\mev/\cunit^{2}$.
The $\Lambda$ signal region ($\left|M_{p\pi^-}-M_{\Lambda}\right|<10~\mev/\cunit^{2}$, red solid box) and the sideband regions ($15<\left|M_{p\pi^-}-M_{\Lambda}\right|<25~\mev/\cunit^{2}$, blue dashed boxes) are indicated.
(Right) The missing mass $\mathrm{MM}(\pi^-pp)$ for events in the $\Lambda$ signal region (black points) and in the combined $\Lambda$ sidebands (hatched histogram).
The vertical dashed lines indicate the applied $\left|\mathrm{MM}(\pi^-pp)-M_{K^{0}}\right|<250~\mev/\cunit^{2}$ window.
}
  \label{fig:MM_IM}
\end{figure}

Figure~\ref{fig:MM_IM} shows the resulting $M_{p\pi^-}$ and ${\mathrm{MM}}(\pi^-pp)$ distributions after all selections.
This event sample provides a relatively clean selection of $\gamma d\to K^{0}\Lambda p$ candidates.

\subsection{Kinematic fit}
A kinematic fit was applied to the $\gamma d \to K^{0}\Lambda p$ candidate events to refine the four-momenta used in the subsequent analysis.
Experimentally, the final state was reconstructed as the $\gamma d \to \pi^- p p X$ reaction, where the incident photon energy and the momentum vectors of the charged particles ($\pi^-$ and two protons) were measured, while the kinematic variables for the recoil $K^0$ were treated to be unmeasured.
The fit was performed with the \texttt{KinFitter} package~\cite{KinFitter}.

The following constraints were imposed: (i) four-momentum conservation for the $\gamma d \to \pi^- p p X$ reaction with the missing particle mass constrained to $M_X=M_{K^{0}}$, and (ii) the invariant-mass constraint $M_{p\pi^-}=M_{\Lambda}$.
The fit performance was validated on data using $\mathcal{P}(\chi^{2})$ and the pull distributions of the fitted variables.
Events with $\mathcal{P}(\chi^{2})<0.01$ were rejected.
The improvement of the $\Lambda p$ invariant-mass resolution was evaluated using MC samples.
The kinematic fit improved the $M_{\Lambda p}$ resolution from $30$ to $11~\mev/\cunit^{2}$, which was essential for a near-threshold search with limited statistics.

\subsection{Observed \texorpdfstring{$(M_{\Lambda p}, q)$}{(M\_Lambda\_p, q)} distribution}
\begin{figure*}[t]
  \centering \includegraphics[width=120mm]{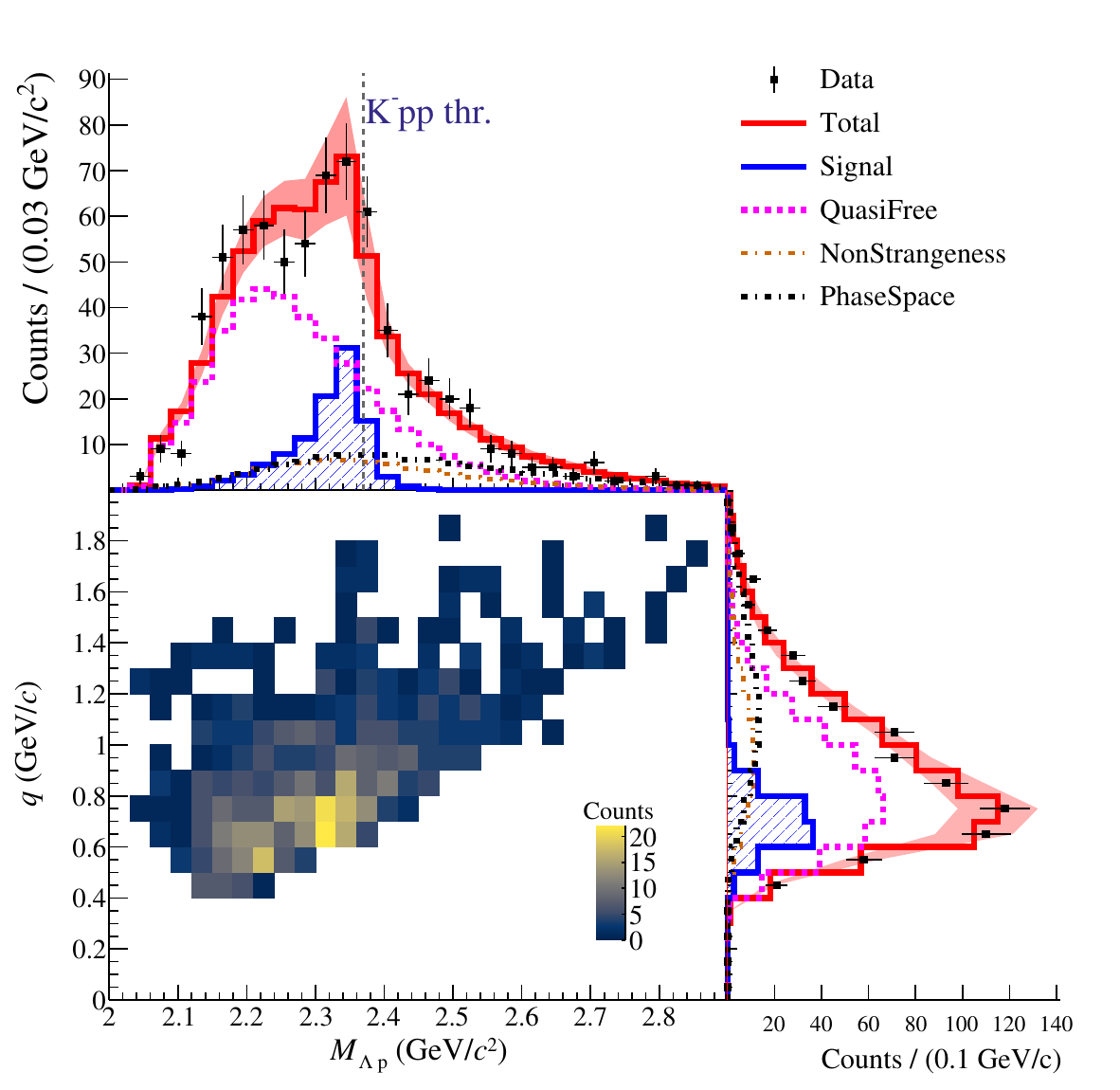}
  \caption{The $(M_{\Lambda p}, q)$ distribution with the superposed result of the template fit.
  The data are compared with the best-fit sum (Total) and the individual components: signal (Sig), quasi-free (QF), non-strangeness (NS), and phase-space (PS).
  The shaded band around the total fit curve represents the fit uncertainty.
  The top panel shows the $M_{\Lambda p}$ projection, the center panel the two-dimensional $(M_{\Lambda p}, q)$ distribution, and the right panel the $q$ projection.
  The vertical dashed line in the upper panel indicates the $K^-pp$ mass threshold.}
  \label{fig:fit2d}
\end{figure*}

To study the reaction dynamics, we examine the two-dimensional distribution of the invariant mass $M_{\Lambda p}$ and the momentum transfer $q$, measured after the kinematic fit.
As required by four-momentum conservation, $q$ is equivalently defined as $q \equiv |\vec{p}_{\gamma}-\vec{p}_{K^{0}}| = |\vec{p}_{\Lambda p}|$.
The $(M_{\Lambda p}, q)$ distribution of the selected $\gamma d \to K^{0}\Lambda p$ candidates is shown in the center panel of Fig.~\ref{fig:fit2d}.
In the two-dimensional distribution, a localized enhancement is observed below the $K^-pp$ mass threshold at low momentum transfer.
In the one-dimensional $M_{\Lambda p}$ projection (top panel), a bump structure is also seen below the threshold.
Similarly, in the one-dimensional $q$ projection (right panel), the distribution peaks around $q \simeq 0.6$--$0.8~\gev/\cunit$ and gradually decreases towards higher momentum transfer.
These features are expected neither from the quasi-free $K^{0}\Lambda$ photoproduction, which populates lower $M_{\Lambda p}$ and broader $q$ regions, nor from a pure phase-space distribution.

\section{Two-dimensional template fit}
To quantify the observed structure, we perform an extended maximum-likelihood fit to the binned two-dimensional histogram.
We fit the distribution in the range $2.0 < M_{\Lambda p} < 2.9~\gev/\cunit^{2}$ and $0 < q < 2.0~\gev/\cunit$, using bin widths of $\Delta M_{\Lambda p}=0.03~\gev/\cunit^{2}$ and $\Delta q=0.1~\gev/\cunit$.
The likelihood is constructed as a product of Poisson probabilities for each bin $(i,j)$, $P(N_{ij}|\mu_{ij})$, where $N_{ij}$ and $\mu_{ij}$ are the observed and expected counts, respectively.
The negative log-likelihood (up to an additive constant) is
\begin{equation}
  \mathrm{NLL} = \sum_{i,j}\left(\mu_{ij} - N_{ij}\ln \mu_{ij}\right).
\end{equation}
The expected yield in each bin is modeled as a linear combination of normalized templates,
\begin{equation}
  \mu_{ij} = \sum_{k} N_k\, T_{k,ij},
\end{equation}
where $T_{k,ij}$ is the content of the $k$-th template normalized to unity over the fit range and $N_k$ is its yield.
All yields are constrained to be non-negative.

The total fit model is constructed using four template components:
(i) quasi-free $\gamma n \to K^{0}\Lambda$ production with a spectator proton (QF),
(ii) a pure three-body phase-space process for $\gamma d \to K^{0}\Lambda p$ (PS),
(iii) a non-strange background estimated from the $\Lambda$ sidebands (NS), and
(iv) a signal component representing a possible $\bar{K}NN$ formation (Sig).
The QF and PS templates were obtained from MC samples processed through the full Geant4-based detector simulation and the same reconstruction and event-selection chain as the data, including the kinematic fit.
Because the templates incorporate the full detector acceptance and kinematic correlations in the $(M_{\Lambda p},q)$ plane, no separate acceptance correction was applied to the data.
We also examined a dedicated MC template for a two-step rescattering process, $\gamma p \to K^{0}\Sigma^{+}$ followed by $\Sigma^{+} n \to \Lambda p$.
Owing to its characteristic $(M_{\Lambda p}, q)$ correlation, this component does not describe the localized enhancement observed in the data, and its fitted yield is consistent with zero.
We therefore conclude that this rescattering process appears insufficient to explain the observed structure and do not include it in the nominal model.

The QF sample was generated based on the existing experimental data and Fermi momentum distributions.
The $\gamma n \to K^{0}\Lambda$ production was weighted to reproduce the center-of-mass energy $W$ and angular dependences reported by CLAS for the $\gamma d \to K^{0}\Lambda(p)$ reaction~\cite{Compton2017_PRC96_065201}, and the bound-nucleon motion is modeled using the AV18 $NN$ interaction~\cite{Wiringa1995_PRC51_38}.
For the PS component, events were generated according to a pure three-body phase-space distribution.
The NS and signal templates described below were constructed by reweighting the PS template, which ensures that these acceptance effects are consistently preserved for all components.
The NS component represents events not associated with a true $\Lambda\to p\pi^-$ decay, as estimated from the $\Lambda$-sideband regions; it can include combinatorial pairs and other non-$\Lambda$ sources.
Its shape is constrained using events in the $\Lambda$ sideband regions shown in Fig.~\ref{fig:MM_IM}.
Due to limited sideband statistics, we construct a smooth NS template by reweighting the PS template with a Gaussian factor in $q$:
\begin{equation}
  T_{\mathrm{NS}}(M_{\Lambda p},q)
  = C_{\mathrm{NS}}
  \exp\!\left(-\frac{q^2}{Q_{\mathrm{NS}}^2}\right)\,
  T_{\mathrm{PS}}(M_{\Lambda p},q),
  \label{eq:NSmodel}
\end{equation}
where $T_{\mathrm{PS}}$ is the normalized PS template after the full analysis chain and $C_{\mathrm{NS}}$ is chosen such that $T_{\mathrm{NS}}$ is normalized to unity within the fit range.
The scale parameter $Q_{\mathrm{NS}}$ is determined by binned two-dimensional fitting using a single template $T_{\mathrm{NS}}$ to the sideband data as a function of $q$.
In the final fit to the $\Lambda$ signal region, the NS yield and $Q_{\mathrm{NS}}$ are fixed at the values determined from the sideband study.
The sideband events are dominated by multi-pion and combinatorial backgrounds that do not contain a real $\Lambda$ decay.
Such non-strange contributions are expected to produce a smooth, phase-space-like distribution in the $(M_{\Lambda p},q)$ plane after the same exclusivity and kinematic-fit requirements are imposed.
No signal-like localized structure is observed in the sideband distribution in the low-$q$, subthreshold region.
The stability of the fit against variations of the NS treatment is discussed below.

We model the signal as a Voigt profile in $M_{\Lambda p}$ (a Breit--Wigner convolved with a Gaussian resolution) multiplied by a Gaussian form factor in $q$.
The signal template is built by reweighting the PS template with the assumed line shape:
\begin{equation}
\begin{aligned}
T_{\mathrm{Sig}}(M_{\Lambda p}, q)
&= C_{\mathrm{Sig}}\,
V(M_{\Lambda p}; M, \Gamma, \sigma_M)\,
\exp\!\left(-\frac{q^2}{Q^2}\right)\,
T_{\mathrm{PS}}(M_{\Lambda p}, q),
\end{aligned}
\label{eq:Sigmodel}
\end{equation}
where $C_{\mathrm{Sig}}$ is chosen such that $T_{\mathrm{Sig}}$ is normalized to unity within the fit region.
$V(M_{\Lambda p}; M, \Gamma, \sigma_M)$ is a Voigt profile, given by the convolution of a non-relativistic Breit--Wigner line shape proportional to $1/[(M_{\Lambda p}-M)^2+\Gamma^2/4]$, with a Gaussian resolution function normalized to unity.
Here, $\Gamma$ is the full width at half maximum of the Breit--Wigner distribution, and
$\sigma_M = 11~\mev/\cunit^2$ is fixed to the effective mass resolution after the kinematic fit in the relevant region.
The free parameters are $M$, $\Gamma$, and $Q$ in the fit.
We perform fits both without and with the signal component and compare their likelihood values.

\section{Results and discussion}
\subsection{Fit results}
Figure~\ref{fig:fit2d} shows the best-fit model and the individual components.
The projections onto $M_{\Lambda p}$ (top) and $q$ (right) compare the data with the best-fit total and individual components.
A fit excluding the signal component fails to reproduce the below-threshold population in the low-$q$ region, whereas including the signal component substantially improves the description.
The fit results are shown in Table~\ref{tab:fitparams}.

\begin{table}[!ht]
\centering
\caption{Fit results including the signal component. The NS yield and $Q_{\mathrm{NS}}$ are fixed from the sideband study.}
\label{tab:fitparams}
\begin{tabular}{@{}lcc@{}}
\toprule
Parameter & Value & Stat. Unc. \\
\midrule
$N_\mathrm{QF}$   & 406 & $\pm 48$ \\
$N_\mathrm{PS}$   & 111 & $\pm 38$ \\
$N_\mathrm{Sig}$  & 102 & $\pm 30$ \\
$N_\mathrm{NS}$   & 78  & - \\
$Q_\mathrm{NS}$ (GeV/$c$) & 1.07 & - \\
$M$ (GeV/$c^2$)      & 2.354 & $\pm 0.011$ \\
$\Gamma$ (GeV/$c^2$)           & 0.055 & $\pm 0.023$ \\
$Q$ (GeV/$c$)  & 0.350 & $\pm 0.041$ \\
\bottomrule
\end{tabular}
\end{table}

The signal parameters yield a mass of $M = 2.354 \pm 0.011\,(\mathrm{stat.})~\gev/\cunit^2$ and a width of $\Gamma = 0.055 \pm 0.023\,(\mathrm{stat.})~\gev/\cunit^2$.
These should be interpreted as effective shape parameters characterizing the localized enhancement in this specific channel.
The extracted parameters remain stable even when the Breit--Wigner line shape is replaced by a form that suppresses the spectral intensity above the $K^-pp$ mass threshold, where the $\bar{K}NN$ decay channel opens.
Since the observed line shape may be distorted by interference with non-resonant amplitudes, the values are not directly comparable across different reaction channels.

The fitted yields indicate that quasi-free $\gamma n \to K^{0}\Lambda$ production accounts for approximately 60\% of the events in the fit region.
The fitted form-factor scale $Q = 0.350 \pm 0.041~\gev/\cunit$ quantifies the concentration at low momentum transfer within the adopted fit model.
The fit results are robust against the treatment of the NS component.
When the fixed NS yield and $Q_{\mathrm{NS}}$ are independently varied within their statistical uncertainties obtained from the sideband study, the changes in the resulting signal parameters are negligible.
As a more conservative cross-check, the fit was repeated with the NS component omitted entirely.
In this case, the broad sideband-like contribution is absorbed mainly by the PS component, whereas the localized enhancement remains assigned to the signal component and the fitted values of $M$, $\Gamma$, and $Q$ remain consistent with the nominal results.

\subsection{Statistical significance and validation}
To evaluate the significance of the signal component, we use the likelihood-ratio test statistic (TS)
$\mathrm{TS}_{\mathrm{obs}}=2(\mathrm{NLL}_{\mathrm{b}} - \mathrm{NLL}_{\mathrm{s+b}}) = 64.9$,
where $\mathrm{NLL}_{\mathrm{b}}$ and $\mathrm{NLL}_{\mathrm{s+b}}$ are the negative log-likelihood values for the background-only and signal-plus-background hypotheses, respectively.
In the signal hypothesis, we profile over the signal yield and three signal-shape parameters $(M,\,\Gamma,\,Q)$, while imposing the physical constraint $N_\mathrm{Sig}\ge 0$.
Thus, the signal hypothesis adds four parameters relative to the background-only hypothesis.
Because the signal yield is restricted to the physical region and the null hypothesis lies on the boundary at $N_\mathrm{Sig}=0$, the regular Wilks theorem is not directly applicable.
We therefore approximate the background-only distribution of TS by a Chernoff-type mixture distribution~\cite{Chernoff1954,Cowan2011},
\begin{equation}
f(\mathrm{TS})=\tfrac{1}{2}\delta(\mathrm{TS})+\tfrac{1}{2}\chi^{2}_{k}(\mathrm{TS})\quad (k=4),
\end{equation}
where $k=4$ corresponds to the signal yield and the three profiled signal-shape parameters.
The $\delta(\mathrm{TS})$ term arises from background-only pseudo-experiments for which the best-fit signal yield is at the boundary, $N_\mathrm{Sig}=0$, giving $\mathrm{TS}=0$.
We compute the local $p$-value as $p_\mathrm{local}=\tfrac{1}{2}\,P(X_4\ge \mathrm{TS}_\mathrm{obs})$, where $X_4\sim\chi^2_4$ denotes a chi-square random variable with four degrees of freedom.
This corresponds to a local significance of $7.3\sigma$.
The asymptotic test-statistic distribution was validated with $10^{5}$ background-only pseudo-experiments using the same fit configuration.
Over the TS range covered by the pseudo-experiments, the cumulative tail probability is well reproduced by the Chernoff-type expectation with $k=4$ and is clearly inconsistent with the $k=1$ boundary case; no pseudo-experiment yielded $\mathrm{TS}\ge \mathrm{TS}_{\rm obs}$.

The global significance may in principle be reduced by a look-elsewhere effect (LEE) associated with the signal-parameter scan.
However, the search region in the present analysis is guided by the prior observation of the $\bar{K}NN$ state in the J-PARC E15 experiment~\cite{Yamaga2020}, which constrains the physically motivated mass range.
For the nominal scan range $2.25 < M < 2.37~\gev/\cunit^2$, the ratio of the cumulative tail probability obtained from the pseudo-experiments to the asymptotic prediction is $1.04$ over the tested TS range.
Even when the scan range is extended down to the $\Lambda p$ threshold ($2.05~\gev/\cunit^2$), the ratio increases only to $1.52$.
Since the asymptotic formula is validated by the pseudo-experiments up to the largest TS values they probe, and the LEE correction factor remains modest even for the conservative scan range, the look-elsewhere effect does not substantially diminish the observed signal significance.

\subsection{Systematic uncertainties}
Systematic uncertainties on the fitted signal parameters are evaluated by repeating the full analysis with alternative selections and modeling choices.
For each source, we prepare samples by varying the corresponding cut parameter within a range motivated by the detector resolution and/or analysis stability.
In particular, for the $\Lambda$ flight-length requirement and the $\pi^-p$ DCA requirement, the nominal thresholds are shifted by $\pm0.5\sigma_v$ and $\pm1.0\sigma_v$, respectively, where $\sigma_v$ denotes the effective vertex resolution.

In addition, the systematic uncertainty associated with the Fermi-momentum distribution in the deuteron is evaluated by regenerating the QF MC template using the Paris~\cite{Lacombe1980Paris} and CD-Bonn~\cite{Machleidt2001CDBonn} deuteron wave-function models and repeating the full analysis.
Since the signal template is constructed by reweighting the phase-space template with the assumed line shape, this variation enters only through the QF background modeling.

From the resulting set of fitted values $\{x_i\}$ for a parameter $x$, we compute the mean $\bar{x}$ and the standard deviation $\sigma$, and interpret $[\bar{x}-\sigma,\ \bar{x}+\sigma]$ as the systematic variation band.
The reported central value is kept as the nominal one ($x_0$), and the asymmetric systematic uncertainties are assigned as
\begin{align}
\delta_\mathrm{sys}^{+} &= \max\!\left(0,\ (\bar{x}+\sigma)-x_0\right), \\
\delta_\mathrm{sys}^{-} &= \max\!\left(0,\ x_0-(\bar{x}-\sigma)\right).
\end{align}
Individual contributions are combined in quadrature separately for the positive and negative sides.
We also investigated the dependence on the histogram binning; the variations were smaller than the statistical uncertainties and strongly correlated with statistical fluctuations, and therefore are not quoted as an independent systematic uncertainty.
The estimated systematic uncertainties are shown in Table~\ref{tab:syst_summary}.

\begin{table}[!ht]
\centering
\caption{Summary of systematic uncertainties on the fitted signal parameters.}
\label{tab:syst_summary}
\small
\setlength{\tabcolsep}{4pt}
\renewcommand{\arraystretch}{1.05}
\begin{tabular}{@{}lccc@{}}
\toprule
Source &
\multicolumn{1}{c}{\shortstack{$M$\\[-0.2ex]\scriptsize (MeV/$c^2$)}} &
\multicolumn{1}{c}{\shortstack{$\Gamma$\\[-0.2ex]\scriptsize (MeV/$c^2$)}} &
\multicolumn{1}{c}{\shortstack{$Q$\\[-0.2ex]\scriptsize (MeV/$c$)}} \\
\midrule
Mass cut             & $+0,\,-4$  & $+1,\,-7$  & $+5,\,-9$   \\
Flight-length cut    & $+7,\,-1$  & $+12,\,-4$ & $+10,\,-2$  \\
DCA cut              & $+4,\,0$   & $+21,\,0$  & $+19,\,0$   \\
PID cut              & $+0,\,-2$  & $+1,\,-3$  & $+3,\,-1$   \\
Fermi-momentum model & $+3,\,-1$  & $+14,\,-2$ & $+25,\,-3$  \\
Kinematic fit        & $+3,\,-1$  & $+13,\,-1$ & $+9,\,0$    \\
\midrule
Total                & $+9,\,-5$  & $+31,\,-9$ & $+35,\,-10$ \\
\bottomrule
\end{tabular}
\end{table}

The largest shifts are induced by the flight-length cut for $M$, the DCA cut for $\Gamma$, and the Fermi-momentum model for $Q$.
The stability against realistic variations of the deuteron wave function indicates that the observed enhancement is not an artifact of a specific Fermi-momentum model.
The effective shape parameters including both statistical and systematic uncertainties are
$M = 2.354 \pm 0.011\,(\mathrm{stat.})^{+0.009}_{-0.005}\,(\mathrm{syst.})~\gev/\cunit^{2}$,
$\Gamma = 0.055 \pm 0.023\,(\mathrm{stat.})^{+0.031}_{-0.009}\,(\mathrm{syst.})~\gev/\cunit^{2}$, and
$Q = 0.350 \pm 0.041\,(\mathrm{stat.})^{+0.035}_{-0.010}\,(\mathrm{syst.})~\gev/\cunit$.

\subsection{Comparison with kaon-induced measurements}
We limit our comparison to the J-PARC E15 experiment, which remains the most relevant reference for the $\bar{K}NN$ quasi-bound state among previous experimental searches.
Other experimental candidates for the $\bar{K}NN$ state are still under active debate due to alternative interpretations, and a direct comparison of peak positions across different reaction channels is not straightforward.
The E15 experiment observed a $\bar{K}NN$ candidate in the $K^{-}\,{}^{3}\mathrm{He} \to \Lambda p n$ reaction~\cite{Yamaga2020}, reporting
$M = 2.328 \pm 0.003\,(\mathrm{stat.})^{+0.004}_{-0.003}\,(\mathrm{syst.})~\gev/\cunit^{2}$,
$\Gamma = 0.100 \pm 0.007\,(\mathrm{stat.})^{+0.019}_{-0.009}\,(\mathrm{syst.})~\gev/\cunit^2$, and
$Q_K = 0.383 \pm 0.011\,(\mathrm{stat.})^{+0.004}_{-0.001}\,(\mathrm{syst.})~\gev/\cunit$.
Our effective mass lies closer to the $K^-pp$ threshold than the E15 value, while the fitted width is consistent with the E15 value within the large uncertainties.
Furthermore, the fitted form-factor scale $Q$ is consistent with the E15 value of $Q_K$.
However, a direct comparison of the line-shape parameters is complicated by reaction-dependent distortions~\cite{Sekihara2016PTEP} and possible interference effects.
Furthermore, the observed enhancement lies more than $0.2~\gev/\cunit^2$ above both the $\Sigma N$ threshold ($\sim\!2.13~\gev/\cunit^2$) and the $\Lambda p$ threshold ($\sim\!2.05~\gev/\cunit^2$), so a simple threshold-cusp explanation alone appears insufficient to account for the localized structure observed in both $M_{\Lambda p}$ and $q$.
More complex coupled-channel effects may still play a role (e.g., those associated with the $\Lambda^{*}N$ or $\Sigma^{*}N$ thresholds). These possibilities cannot be fully excluded; however, the localized nature of the structure is consistent with a quasi-bound-state interpretation.
Combined analysis of results from both hadron- and photo-induced reactions is therefore essential to achieve a deeper understanding of kaonic nuclei.

\subsection{Forward \texorpdfstring{$K^0$}{K0} emission in the signal region}

\begin{figure}[ht]
  \centering
  \includegraphics[width=0.70\linewidth]{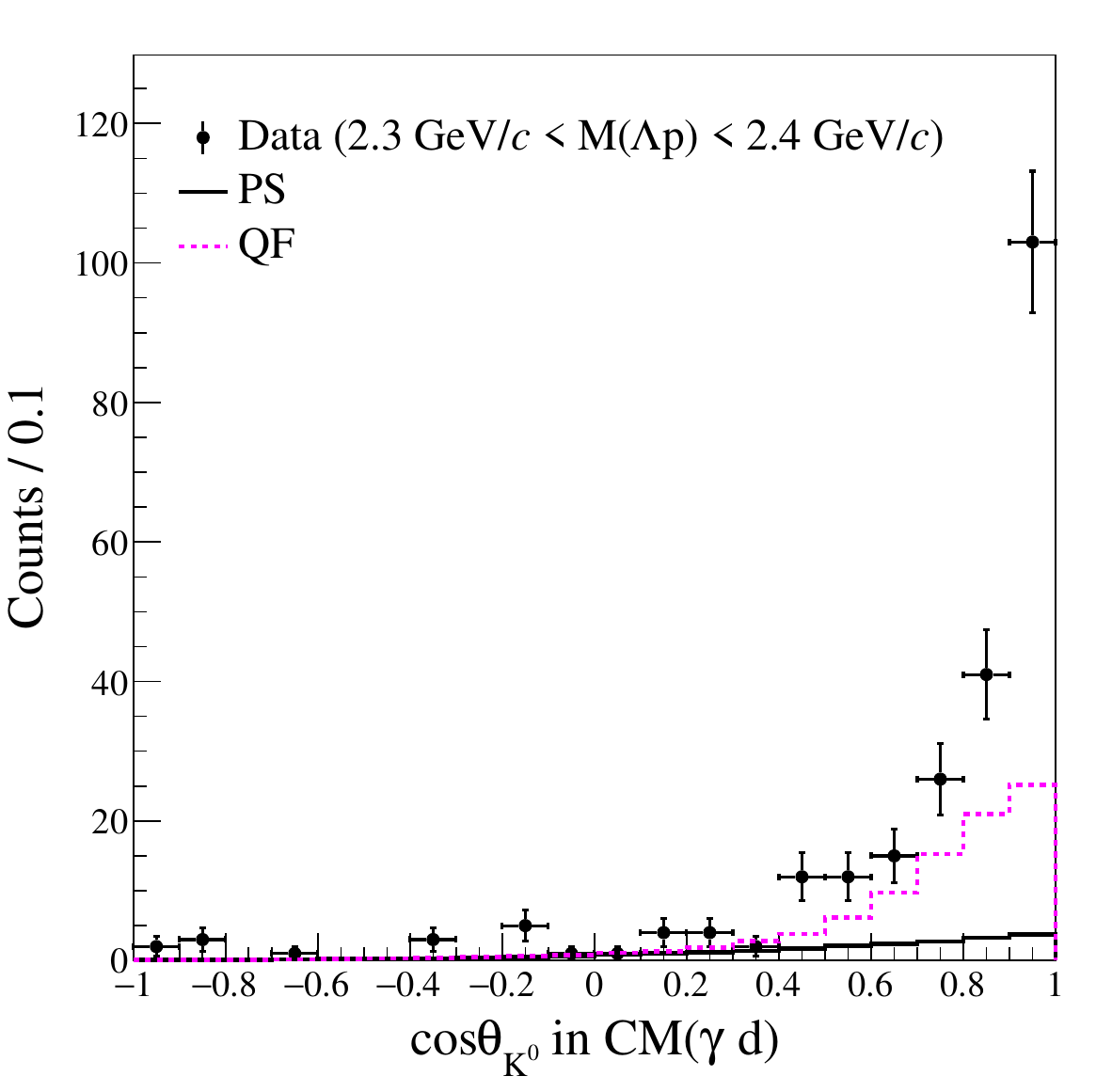}
  \caption{$\cos\theta^{\rm CM}_{K^0}$ distribution in the $\gamma d$ center-of-mass frame for events with $2.3<M_{\Lambda p}<2.4~\gev/\cunit^2$.
  Data (points) are compared with the quasi-free $\gamma n \to K^{0}\Lambda$ MC (QF, dashed) and the pure three-body phase-space $\gamma d \to K^{0}\Lambda p$ MC (PS, solid), both normalized to their respective fitted yields.
  A prominent forward peak at $\cos\theta^{\rm CM}_{K^0} \gtrsim 0.8$, beyond the gradual acceptance-induced rise in the PS distribution, is consistent with both the kinematics of quasi-bound state formation and a $t$-channel $K^*$ exchange production mechanism.}
  \label{fig:costheta_K0}
\end{figure}
Figure~\ref{fig:costheta_K0} shows the $\cos\theta^{\rm CM}_{K^0}$ distribution in the $\gamma d$ center-of-mass frame for events in the signal region ($2.3<M_{\Lambda p}<2.4~\gev/\cunit^2$).
For comparison, the MC distributions for the quasi-free $\gamma n \to K^{0}\Lambda$ process (QF, dashed) and the pure three-body phase-space $\gamma d \to K^{0}\Lambda p$ process (PS, solid) are overlaid after the full analysis chain;
each histogram is normalized to the yield obtained from the two-dimensional fit.
While the PS distribution exhibits a gentle forward rise due to the detector acceptance, the data show a much more prominent forward peaking at $\cos\theta^{\rm CM}_{K^0} \gtrsim 0.8$.
The forward emission of the $K^0$ is kinematically equivalent to a low three-momentum transfer $q$ to the $\Lambda p$ system.
Therefore, this forward concentration is consistent both with the low-momentum-transfer condition expected for quasi-bound-state formation and with a production mechanism dominated by $t$-channel $K^*$ exchange.
The present data, however, do not yet disentangle these two effects, and additional polarization observables will be important for clarifying the production dynamics.

\section{Summary}
We performed an exclusive measurement of the $\gamma d \to K^{0}\Lambda p$ reaction at $E_{\gamma}=1.3$--$2.4$~GeV using the LEPS2 solenoid spectrometer to search for a $\bar{K}NN$ quasi-bound state.
The analysis of the two-dimensional $(M_{\Lambda p}, q)$ distribution reveals a significant enhancement below the $K^-pp$ mass threshold at low momentum transfer.
The local statistical significance of the signal component is $7.3\sigma$.
The observed structure is characterized by a mass $M = 2.354 \pm 0.011(\mathrm{stat.})^{+0.009}_{-0.005}(\mathrm{syst.})~\gev/\cunit^2$, a width $\Gamma = 0.055 \pm 0.023(\mathrm{stat.})^{+0.031}_{-0.009}(\mathrm{syst.})~\gev/\cunit^2$, and a Gaussian form-factor scale $Q = 0.350 \pm 0.041(\mathrm{stat.})^{+0.035}_{-0.010}(\mathrm{syst.})~\gev/\cunit$.
These results provide the first evidence for a $\KNN$ quasi-bound state in photoproduction.
The consistency of the form-factor scale with the J-PARC E15 result suggests a common physical origin, while the difference in the effective mass highlights the reaction dependence of the observed line shape.
The prominent forward-peaking of the $K^0$ emission (Fig.~\ref{fig:costheta_K0}), which kinematically corresponds to the low-momentum-transfer region, is consistent with both the formation condition of the quasi-bound state and a $t$-channel $K^*$ exchange production mechanism.
This work demonstrates that photoproduction provides a complementary and powerful probe of kaonic nuclear states.

Further analysis, including beam asymmetry measurements exploiting the linear polarization of the LEPS2 photon beam, as well as searches for isospin-partner channels and studies of other decay modes, will be essential to unambiguously identify the production mechanism and determine the quantum numbers of the $\bar{K}NN$ state.

\section*{Acknowledgments}
The authors thank the SPring-8 accelerator staff for stable beam operation.
The authors also thank Y.~Yanai, R.~Yamamoto, Y.~Kasamatsu, H.~Ikuno, T.~Noritake, H.~Goto, M.~Tsuruta, K.~Nishi, T.~Nobata, H.~Saito, R.~Shirai, Y.~Furuta, K.~Nakamura, T.~Shibata, R.~Mizuta, Y.~Narita, E.~Umezaki, D.~Nakamura, K.~Miki, S.~Shibuya, E.~A.~Strokovsky, D.~Krivenkov, A.~Averyanov, H.~Kawai, M.~Kubo, M.~Tabata, Triloki, R.~Rai, K.~Nanbu, I.~Nagasawa, and T.~N.~Takahashi for their helpful cooperation during the preparation stage of the experiment.
This work was supported in part by JSPS KAKENHI Grant Numbers 15H00839, 16K13806, 17H02892, 18H05402, 20H01933, 21H00114, 21H01110, 21H01115, 21H04986, 22H00124, 23H01201, and 24H00232; by the National Science and Technology Council of Taiwan (R.O.C.); and in part by the National Research Foundation of Korea under Grant Number 2020R1A3B2079993.
\section*{Declaration of Generative AI and AI-assisted technologies in the writing process}
During the preparation of this work, the authors used ChatGPT (OpenAI) in order to improve language and readability. After using this tool/service, the authors reviewed and edited the content as needed and take full responsibility for the content of the publication.

\bibliographystyle{elsarticle-num}
\bibliography{references}

\end{document}